\documentstyle[12pt]{article}

\newcommand{\be}{\begin{equation}}
\newcommand{\ee}{\end{equation}}
\newcommand{\ba}{\begin{eqnarray}}
\newcommand{\ea}{\end{eqnarray}}
\newcommand{\baa}{\begin{eqnarray*}}
\newcommand{\eaa}{\end{eqnarray*}}
\newcommand{\bb}{\begin{thebibliography}}
\newcommand{\eb}{\end{thebibliography}}
\newcommand{\ci}[1]{\cite{#1}}

\newcommand{\lab}[1]{\label{#1}}
\newcommand{\re}[1]{(\ref{#1})}

\newcounter{hran}
\newcommand{\eqalph}{\setcounter{hran}{\value{equation}}
\refstepcounter{hran}\setcounter{equation}{0}
\renewcommand{\theequation}{\thehran\alph{equation}}}

\newcommand{\eqarab}{\setcounter{equation}{\value{hran}}
\renewcommand{\theequation}{\arabic{equation}}}

\newcommand{\ah}{\ifmmode\hat{a}\else$\hat{a}$\fi}
\newcommand{\ahd}{\ifmmode\hat{a}^\dagger\else$\hat{a}^\dagger$\fi}
\newcommand{\ad}{\ifmmode{a^\dagger}\else$a^\dagger$\fi}
\newcommand{\bx}{\ifmmode{\mbox{\bf x}}\else{\bf x}\fi}
\newcommand{\by}{\ifmmode{\mbox{\bf y}}\else{\bf y}\fi}
\newcommand{\dl}{\ifmmode\delta\else$\delta$\fi}
\newcommand{\Dl}{\ifmmode\Delta\else$\Delta$\fi}
\newcommand{\al}{\ifmmode\alpha\else$\alpha$\fi}
\newcommand{\ph}{\ifmmode\hat\varphi\else$\hat\varphi$\fi}
\newcommand{\vp}{\ifmmode\varphi\else$\varphi$\fi}
\newcommand{\pl}{\ifmmode\partial\else$\partial$\fi}
\newcommand{\kp}{\ifmmode\kappa\else$\kappa$\fi}
\newcommand{\tet}{\ifmmode\theta\else$\theta$\fi}
\newcommand{\eps}{\ifmmode\varepsilon\else$\varepsilon$\fi}
\newcommand{\derl}[1]{\ifmmode\frac{\stackrel{\rightarrow}{\pl}}{\pl #1}
        \else$\frac{\stackrel{\rightarrow}{\pl}}{\pl #1}$\fi}
\newcommand{\derr}[1]{\ifmmode\frac{\stackrel{\leftarrow}{\pl}}{\pl #1}
        \else$\frac{\stackrel{\leftarrow}{\pl}}{\pl #1}$\fi}
\newcommand{\der}[1]{\ifmmode\frac{\pl}{\pl #1}
	\else$\frac{\pl}{\pl #1}$\fi}
\newcommand{\dertwo}[2]{\ifmmode\frac{\pl #1}{\pl #2}
	\else$\frac{\pl #1}{\pl #2}$\fi}
\newcommand{\N}{\ifmmode\mbox{\rm N}\else {\rm N}\fi}
\newcommand{\T}{\ifmmode\mbox{\rm T}\else {\rm T}\fi}
\newcommand{\Sym}{\ifmmode\mbox{\rm Sym}\else {\rm Sym}\fi}
\newcommand{\PPr}{\ifmmode{\stackrel{\leftarrow}{\cal P}}
	\else $\stackrel{\leftarrow}{\cal P}$\fi}

\newcommand{\Tr}{\mbox{Tr}}

\title{q-Functional Wick's theorems for particles
with exotic statistics}
\author{
K.N.Ilinski $^{1,2,3}$
\thanks{E-mail: kni@th.ph.bham.ac.uk}, \
G.V.Kalinin $^{1}$
\thanks{E-mail: kalinin@snoopy.phys.spbu.ru} \
and A.S.Stepanenko $^{1,2,4}$
\thanks{E-mail: ass@th.ph.bham.ac.uk}\\ [0.2cm]
{\small\it $^{1}$ IPhys Group, CAPE, St.-Petersburg State
University,}\\
{\small\it 14-th line of V.I. 29, St-Petersburg, 199178, Russian
Federation}
\\ [0.2cm]
{\small\it $^{2}$ School of Physics and Space Research,
University of Birmingham,} \\
{\small\it Edgbaston B15 2TT, Birmingham, United Kingdom}
\\ [0.2cm]
{\small\it $^{3}$ Theoretical Department, Institute of Spectroscopy,
Troitsk,\ \ \ \ \ \ \ \ \ \ \ } \\
{\small\it  \ \ \   Moscow region, 142092, Russian Federation }
\\ [0.2cm]
{\small\it $^{4}$ Theoretical Department, St.-Petersburg Nuclear Physics
Institute,}\\
{\small\it  Gatchina, St.-Petersburg, 188350, Russian Federation}
}
\date{  }

\begin{document}
\baselineskip 16pt
\maketitle

\begin{abstract}
In the paper we begin a description of functional methods of
quantum field theory for systems of interacting $q$-particles. These
particles obey exotic statistics and are the $q$-generalization of the
colored particles  which appear in many problems of condensed matter
physics, magnetism and quantum optics. Motivated by the general
ideas of standard field theory we prove the $q$-functional
analogues of Hori's formulation of Wick's theorems for the different
ordered
$q$-particle creation and annihilation operators. {\it The formulae have
the same formal expressions as fermionic and bosonic ones but differ by
a nature of fields.} This allows us to derive the perturbation series
for the theory and develop analogues of standard
quantum field theory constructions in $q$-functional form.
\end{abstract}

\newpage

\section{Introduction}
Last decade many mathematical structures were deformed and
gained the subscript $q$ in their notations.
In this way remarkable mathematical objects such as non-commutative
geometries~\cite{NG}, quantum groups ($q$-groups)~\cite{QG} and their
representations on quantum vector spaces~\cite{Mn} have arisen.
These objects led to the investigations of the quantum group gauge
theory~\cite{QGGT}, $q$-deformed Schrodinger
equation~\cite{DSE} and classical and quantum dynamics on $q$-deformed
phase spaces~\cite{DD}. In Refs. \cite{Mar,BI} $q$-deformed spaces were
considered as  graded-com\-mu\-ta\-ti\-ve algebras.  On this basis the
classical and quantum dynamics on $q$-deformed spaces were proposed in
close analogy with the case of {\bf Z$_2$}-commutative spaces
(Grassmann
algebras)~\cite{BI}.

After quantization the particles satisfy
the $q$-de\-for\-med commutation relations similar to  the commutation
relations of the colored particles \cite{Ohn}:
\be
\begin{array}{l}
a_k a_j \mp q a_j a_{k} = 0\ ,\quad
a_{k} a_{j}^+ \mp q^{-1} a_{j}^+ a_{k} = 0 \ ,\quad
q = e^{i\alpha}\ ,\quad  1\leq k<j\leq n \ ,
\\
a_{k} a_{k}^+ \mp a_{k}^+ a_{k} = 1
\qquad \mbox{and for $q$-deformed fermions}\quad
[a_k^+]^2 = [a_k]^2 = 0\ .
\end{array}
\label{q-p}
\ee
where the upper sign corresponds to the quantization on $q$-deformed
common space and the lower one corresponds to the quantization on
$q$-deformed Grassmann algebra. These particles were called
$q$-particles
to distinguish them from the famous $q$-oscillators which a number of papers
are devoted to.  These $q$-particles are the only objects of our
consideration.

It is surprising that $q$-particles are interesting not only from
the mathematical point of view but also from the physical one. Here we do
not mean the appearance of $q$-particles in the
parastatistics~\cite{Mar,BI,F}, $q$-extended Supersymmetry~\cite{IU1},
Parasupersymmetry~\cite{IU1,BD} and other similar problems because
they are rather mathematical applications. By physical
applications we mean firstly solid state physics and quantum
optics where $q$-particles appear in a natural way as well as the
theory of magnetics where Paulions originally appeared and play a
central role.

In solid state physics, anyons (particles with exotic braiding
statistics) are important in some attempts to understand the physical
features of planar systems~\cite{Fr}. The main physical interest in
anyon systems is their possible connection with some effects in two
dimensional condensed matter physics in particular in the quantum Hall
effect~\cite{QH} and high temperature superconductivity~\cite{sc} (in
the framework of the investigations of the $t$-$J$ and Hubbard models).

In contrast to the previous example where anyons serve as auxiliary objects
for the construction of one of possible scenarios  there is a
wide field in the quantum nonlinear optics in which $q$-particles are the main
components.  This is a theory of the collective behavior of excitons with
small radius (Frenkel Excitons and Charge-Transfer Excitons (CTE))
\cite{A1}.  The studies investigate possibilities of formation of the
Frenkel biexcitons and the observation of phase transitions in exciton
systems in molecular crystals (Bose- Einstein condensation of excitons
\cite{AT}, structural phase transitions in crystals with high excitonic
concentrations, dielectric-metal phase transition in a system of
interacting CTE \cite{AI} and others). Strictly speaking excitons are
not
particles. They are quasiparticles describing molecular excitations and
are of great  importance in the analysis of nonlinear optical processes
which accompany propagation of high-intensity light fluxes whose
frequencies are in the range of the exciton absorption bands \cite{A2}.
Moreover excitons obey exotic statistics (Pauli statistics) \cite{A0}
coinciding with $q$-particles statistics for $q=-1$.
The general case of $q=e^{i\alpha}$ arises if we try to take
into account phenomenologically some nonlinear effects (such as the
difference in the creation time of molecular excitations for different
types of molecules). This effect can be modeled by the change of the
Paulion commutation relations to those of the $q$-particles using the
method developed in \cite{PS}.

Surprisingly, even
the investigation of the behavior of low dimensional exciton systems is
meaningful. For example the exact
solutions for 1-dimensional Paulion chains \cite{LMS} caused
great advances in the theory of the so called J-aggregates, {\it i.e.} molecular
aggregates with unusually sharp absorption band (\cite{KS} and Refs.
therein). The investigations of exciton systems on interfaces
closely connect with the successes of contemporary technology .
All these show that $q$-particles find deep
applications in modern physical theories and motivate our objective
to derive the appropriate field theoretical technique for them.

The paper is organized as follows. In the next section we motivate the
necessity for the introduction of $q$-ordered objects (such as the
normal product) and introduce useful notation. Then in the sections 3
and 4 we prove Wick's theorems in $q$-functional form for the simple
product and $q$-symmetrical products of $q$-operators ($q$-symmetrical
and $q$-chronological products). This allows us to find in section 5 the
Wick's theorems for the $q$-operator functionals. The last section contains
the conclusions and some remarks.

\section{Motivation and notations}

In this section we motivate the necessity to introducing $q$-objects
(such as N-, T-products of $q$-operators and others) which we study in
the following sections. We also define ``universal" notations following
\cite{V1} that allow us to prove Wick's theorems for the cases of
$q$-operators with internal degree of freedom (spin, colors) in the
unified way.

As in the usual Fermi-Bose situation in $q$-particle physics we are
interested in two kinds of physical quantities. The first one is
generally speaking the class of equilibrium thermodynamical
characteristics calculated via the partition function of the system.
The second one is a set of correlators which are important in
describing the kinetics.

For clarity we consider a system of $q$-particles with creation
(annihilation) operators obeying the commutation relations (\ref{q-p}).
The equilibrium thermodynamics of the model is described by the
statistical operator $\rho=e^{-\beta H}$ and the partition function
$Z=\Tr\rho$. In close analogy with undeformed many-particle
quantum theory there are several approaches for calculating $Z$. The
comparative convenience of each of them is defined by the specific
features of the system in question. The most straightforward way is to
present $\rho$ by the series
$$
\rho = \sum_{n=0}^{\infty} \frac{(-\beta)^n}{n!} H^{n}
$$
and to find matrix elements of the operators $H^n$ in the occupation number
basis $\{|\varphi_{i_1,i_2,\dots,i_m}\rangle\}$ $=
\{a^+_{i_1}a^+_{i_2}{\dots}a^+_{i_m}|0\rangle\}$
and then to try to calculate the sum of all diagonal matrix elements in
this basis. Usually it is difficult enough to calculate the sum in
a closed form.

To escape this problem one can introduce the basis coherent states
corresponding to creation (annihilation) operators
(\ref{q-p})~\cite{IS}.
These states use elements of the graded-commutative algebra $HQ^{0|L}$ (or
$HQ^{L|0}$ in the bosonic case) as parameters.
By proceeding this way we need to calculate matrix elements of the operator in
the Bargmann-Fock representation. This means we need to find a representation of the
operator in N-ordering form in which all creation operators are on the
right-hand side with respect to all annihilation operators
in all monomial terms. This can be done using the Wick's
theorem for $q$-operators in the same way that it was done for
the usual Fermi-Bose statistics~\cite{V1}. The next section is devoted to the
formulation and proof of the theorem.

It is a very natural idea to use the objects of the non-commutative
algebra for the investigations of systems with quantum particles obeying
some exotic statistics. The first example of its useful application is
in fermionic physics.
Moreover all functional methods of quantum bosonic field theory can be
modified for the treatment of quantum fermionic fields using Grassmann
variables instead of the usual complex numbers.
This allows us to hope that the methods can be adopted for particles
with exotic statistics ($q$-statistics) if permutation relations for the
classical analogues are defined according to the rules of the
statistics.
In this way we can obtain a self-consistent scenario
for building the quantum field theory (QFT) of particles
with exotic statistics.

In this paper we consider creation and annihilation operators obeying
the following commutation relations which generalize (\ref{q-p}):
\be
\left.
\begin{array}{l}
{\ah}(\bx)\ah(\by) -\kp q(\bx,\by)\ah(\by)\ah(\bx) = 0\ ,
\\
{\ahd}(\bx)\ahd(\by) -\kp q(\bx,\by)\ahd(\by)\ahd(\bx) = 0\ ,
\\
{\ah}(\bx)\ahd(\by) -\kp q^*(\bx,\by)\ahd(\by)\ah(\bx)
= \dl(\bx,\by)\ ,
\\ {}
[\ah(\bx)]^2 = [\ahd(\bx)]^2 = 0\ ,\quad \mbox{for $q$-fermions}\ ,
\end{array}
\right\}
\lab{qoper}
\ee
This form allows us to consider continuous indices as well as discrete
ones. \bx, \by\ are in general $D$-dimensional vectors (or/and
$D$-dimensional multi-indices for a lattice) which describe external and
internal degree of freedom. The function $\dl(\bx,\by)$ is \dl-function
$\dl(\bx-\by)$ for a continuous space and Kronecker \dl-function for a
lattice or internal color indices. The statistical factor $q(\bx,\by)$
possesses the following property generalized Paulionic and Anyonic
property:
\be
q(\bx,\by) = q^{-1}(\by,\bx) = q^{*}(\by,\bx) \ ,\qquad
q(\bx,\bx) = 1\ .
\lab{propq}
\ee
Finally $\kp$ serves to unify formulae for deformed bosonic and
deformed fermionic cases. As usual it has the form:
\be
\kp=\left\{
\begin{array}{l}
+1\ , \quad \mbox{for $q$-bosons}\\
-1\ , \quad \mbox{for $q$-fermions} \ .
\end{array}
\right.
\lab{kappa}
\ee
Hereafter letters with hats denote operators and those without
hats denote the corresponding classical variables. For operator algebra
\re{qoper} the corresponding classical variables satisfy the
following permutation relations:
\be
\left.
\begin{array}{l}
a(\bx)a(\by) -\kp q(\bx,\by)a(\by)a(\bx) = 0\ ,
\\
{\ad}(\bx)\ad(\by) -\kp q(\bx,\by)\ad(\by)\ad(\bx) = 0\ ,
\\
{a}(\bx)\ad(\by) -\kp q^*(\bx,\by)\ad(\by)a(\bx) = 0\ ,
\\ {}
[a(\bx)]^2 = [\ad(\bx)]^2 = 0\ ,\quad \mbox{for $q$-fermions}\ ,
\end{array}
\right\}
\lab{qclvar}
\ee
Let us introduce ``universal" notations to escape repetitions and
enable us
formulate statements in unified form for different cases. Following
\ci{V1} we collect the creation and annihilation operators into the
single vector ``field" operator:
\be
\ph(x)=
\left(\begin{array}{c}
{\ahd}(\bx)\\{\ah}(\bx)
\end{array}\right)\ ,\quad
x\equiv (s,\bx)\ ,
\lab{univnot}
\ee
such that the additional vector index $s$ indicates a type of operator
(\ahd\ (when $s=1$) or \ah\ (when $s=2$)). It is convenient to represent
quantities by vectors or matrices with respect to the index $s$.

Using this notation the commutation relations \re{qoper} can be
rewritten in the following form:
\be
\left.\begin{array}{l}
{\ph}(x)\ph(y) -\kp Q(x,y)\ph(y)\ph(x) = u(x,y)\ , \\ {} [\ph(x)]^2
= 0\ ,\quad \mbox{for $q$-fermions}\ , \end{array} \right\}
\lab{qphiop}
\ee
where the statistical matrix $Q(x,y)$ is defined by statistical factor
$q(\bx,\by)$:
\eqalph
\be Q(x,y)= \left(
\begin{array}{cc}
q(\bx,\by)&q^*(\bx,\by)
\\
q^*(\bx,\by)&q(\bx,\by)
\end{array}
\right)\ ,
\lab{Q}
\ee
and the quantum deformation parameter matrix is given by the equality:
\be
u(x,y)=\dl(\bx,\by)
\left(
\begin{array}{cc}
0&-\kp\\1&0
\end{array}
\right)\ .
\lab{u}
\ee
\eqarab
The row number of the matrices from \re{Q} and \re{u} corresponds to $s_x$ and
the column number to $s_y$. We can now associate classical
analogues --- classical vector ``field" variables -- with the
corresponding vector ``field" operator. For the classical
variables we get the permutation relations in the following
``universal" form:
\be
\left.
\begin{array}{l}
{\vp}(x)\vp(y) -\kp Q(x,y)\vp(y)\vp(x) = 0\ ,
\\ {}
[\vp(x)]^2 = 0\ ,\quad \mbox{for $q$-fermions}\ ,
\end{array}
\right\}
\lab{qphicl}
\ee
which is defined by the statistical matrix $Q(x,y)$ from \re{Q}.

To formulate Wick's theorems in the functional form we need to
introduce left (\derl{\vp(x)}) and right (\derr{\vp(x)}) functional
derivatives on the algebra \re{qphicl}. We do as was done for the
Grassmann algebra case: to act by the left (right) derivative on an
expression we must bring the corresponding variable to the extremely
left (right) position in accordance with the permutation relations
and cancel it. This leads to the following commutation relations for
the left derivative:
\be
\derl{\vp(x)} \vp(y) -\kp Q^*(x,y) \vp(y) \derl{\vp(x)}
= \dl(x,y)\ ,
\lab{derl}
\ee
where $\dl(x,y)=\dl_{s_x,s_y}\dl(\bx,\by)$.

To construct Hori's functional expressions we will
also need extended algebras which will contain several copies of the
algebra of our vector ``field" variables as sub-algebras. The mutual
permutation relations for the sub-algebras we define to keep
the graded-commutative structure. For example, the pair of fields
$\vp_1(x), \vp_2(x)$ like \re{qphicl} from different sub-algebras
commute with one another as follows:
\be
\vp_1(x)\vp_2(y) -\kp Q(x,y)\vp_2(y)\vp_1(x) = 0\ .
\lab{qphi12}
\ee
We are now ready to proceed with Wick's theorems for the operators
\re{qphiop}.

\section{Wick's theorem for the simple product of $q$-operators}

In this section we formulate and prove the analog of Hori's formula
\cite{Hori} which gives the functional form of the Wick's theorem
for the normal form of a simple product of the creation and annihilation
(or vector ``field") operators. {\it The formal expression does not
depend
on the statistics of the fields.}  It shows that the exotic
statistics can be taken into account by changing the permutation
relations for the classical analogues according to the rules of the
statistics and keeping the formal expression for the Hori's formulae.

As was said above we define the normal form $N(A)$ of the monomial
operator $A$ as an expression where all creation operators
are placed on the left hand side in respect to the annihilation
operators (using permutation relations (\ref{qclvar})). Then the
definition is generalized for an arbitrary polynomial operator by the
linearity. The definition is standard~\cite{V1}.

As usual we can do this in the vector ``field" form (in the ``universal"
notations) where the final expressions only contain the normal contractions
$n(x,y)$. For the pair of ``field" operators the normal product is
defined by the following relation:
\be
{\N}[\ph(x)\ph(y)] = \ph(x)\ph(y) - n(x,y)\ ,
\lab{normprod2}
\ee
For system like \re{qoper} the normal contraction hence has the form:
\be
n(x,y) = \dl(\bx,\by)
\left(
\begin{array}{cc}
0&0\\
1&0
\end{array}
\right)
\lab{pairing}
\ee
We note that the normal contraction has
$\delta$-functional (or $\delta$-symbol) character with respect to the
indices. In the following we will frequently use this fact. More
precisely we will use the fact that the product $\vp(x)\vp(y)$ is
$C$-number function when $n(x,y)\ne0$.

We now prove the following theorem which reduces
the simple operator product to the normal form:

\vspace{0.2cm}
{\bf Theorem 1}
\be
\ph(x_1)\dots\ph(x_n) = \N
\left[\left.
\prod_{i<k}\left(1+\derr{\vp_i}n\derr{\vp_k}\right)
\vp_1(x_1)\dots\vp_n(x_n)\right|_{\dots}\
\right]
\lab{normformr}
\ee
where
\begin{enumerate}
\item the variables $\vp_1,\dots,\vp_n$ form the extended algebra
\re{qphicl} following eq.\re{qphi12},
\item The symbol $|_{\dots}$ means substitution
$\vp_1=\dots=\vp_n=\ph$,
\item Substituting the differential forms into eq.\re{normformr}
in expanded form we have
\be
\der{\vp_i}n\der{\vp_k} = \sum\limits_{x,x^\prime}
\der{\vp_i(x)}n(x,x^\prime)\der{\vp_k(x^\prime)}\ .
\lab{difform}
\ee
(here the symbol $\sum_x$ means summation over discrete variables and
integration over continuous ones).
\end{enumerate}
{\sl Proof.} The strategy of the proof coincides with that
of the proof of the Fermi-Bose Hori's formulae \cite{V1} and is fulfilled
by induction. Indeed, for the particular cases $n=1,2$
the statement can be proved immediately:
\be
\left.\begin{array}{l}
{\ph(x)} = {\N}[\ph(x)]\ ,\\
{\ph(x_1)}\ph(x_2) = {\N}[\ph(x_1)\ph(x_2) + n(x_1,x_2)]\ ,
\end{array}\right\}
\lab{normprod12}
\ee
by the application of the definitions of the normal form and the normal
contraction (\ref{normprod2},\ref{pairing}).

Let us now assume that eq.\re{normformr} is true for any $n\leq N$ and
consider the product of $N+1$ operators $\ph(x_1)\dots\ph(x_{N+1})$.

1. If the field $\ph(x_{N+1})$ contains only the second component
$\ah(\bx_{N+1})$ then
\be
\ph(x_1)\dots\ph(x_{N+1}) = \ph(x_1)\dots\ph(x_N)\ah(\bx_{N+1})
\lab{ah}
\ee
By the inductive assumption the first $N$ multipliers in the RHS of
eq.\re{ah} can be mapped to the normal form by the following reduction
operator $\PPr_N$:
\be
\PPr_N =
\prod_{1\le i<k\le N}\left(1+\derr{\vp_i}n\derr{\vp_k}\right)\ .
\lab{P}
\ee
Then the full expression (\ref{ah}) also takes the normal form. It is
not
difficult to see that due to the structure of normal contraction
(\ref{pairing}) the differential expression (\ref{difform}) in
(\ref{normformr}) does not contain the derivative on
$\ph^{(2)}(x_{N+1})=\ah(\bx_{N+1})$. Hence the terms with derivative
\derr{\vp_{N+1}} do not contribute in this case. So we can substitute
$\PPr_N$ by $\PPr_{N+1}$ what finishes the proof for this case.

2. Let us now consider the case of $\ph(x_{N+1})=\ahd(\bx_{N+1})$.
Using $N$ times the commutation relations (\ref{qphiop}) we obtain the
following relation:
\ba
{\ph}(x_1)\dots\ph(x_{N+1}) &=&
Q \ahd(\bx_{N+1})\ph(x_1)\dots\ph(x_N) +
\nonumber\\
&+& \sum\limits_{k=1}^N Q_k n(x_k,x_{N+1})
[\ph(x_1)\dots\ph(x_N)]_k\ ,
\lab{ahd}
\ea
where $[\ph(x_1)\dots\ph(x_N)]_k$ means product
$\ph(x_1)\dots\ph(x_N)$ without multiplier $\ph(x_k)$ and
the following notations
$$
Q = \kp^N Q(x_1,x_{N+1})\dots Q(x_N,x_{N+1})\ ,
$$
$$
Q_k = \kp^{N-k} Q(x_{k+1},x_{N+1})\dots Q(x_N,x_{N+1})
$$
were introduced. The multipliers $Q$ and $Q_k$ reflect the
$q$-deformation of the statistics and are due to the commutation
relations (\ref{qphiop}). The first summand in the RHS of eq.\re{ahd} can be
rewritten as
\be
\N[Q\ahd(\bx_{N+1})\cdot\PPr_N\vp_1\dots\vp_N|_{\dots}]
\lab{Nahd}
\ee
Let us now note that under the sign of the normal products (\ref{Nahd})
the operator $\ahd(\bx_{N+1})$  can be substituted by the corresponding
classical variable. Then if we move the variable $\ad(\bx_{N+1})$ to the
original (right) position into expression (\ref{Nahd}) we gain the
statistical multiplier $Q^{-1}$ which is exactly cancels the multiplier
$Q$. This is because the reduction operator $\PPr_N$ commutes with
$\ad(\bx_{N+1})$ due to the $\delta$-character of the normal contraction
and the permutation relations for the classical analogues give the same
statistical phases as the quantum ones. So
\be
\N\left[
Q\ahd(\bx_{N+1})\PPr_N\vp_1(x_1)\dots\vp_N(x_N)\Biggr|_{\dots}
\right] =
\N\left[
\PPr_N\vp_1\dots\vp_{N+1}\Biggr|_{\dots}
\right]\ .
\lab{Nahd1}
\ee
Now consider the second summand in RHS of \re{ahd}. By applying
the inductive assumption and substituting normal contraction by the
expression in parentheses we get:
$$
\N\left[\sum\limits_{k=1}^N Q_k
\left(\derr{\vp_k}n\derr{\vp_{N+1}}\vp_k\vp_{N+1}\right)
\PPr_N[\vp(x_1)\dots\vp(x_N)]_k\right]\ .
$$
Product $[\vp(x_1)\dots\vp(x_N)]_k$ does not contain fields $\vp_k,
\vp_{N+1}$. Using arguments similar to those for the derivation of
(\ref{Nahd1}) we can move them to there original position and get
\be
\N\left[\left.\sum\limits_{k=1}^N \derr{\vp_k}n\derr{\vp_{N+1}} \PPr_N
\vp_1\dots\vp_{N+1}\right|_{\dots}\right]\ .
\lab{NPf}
\ee
Collecting together eqs.(\ref{ahd}), (\ref{Nahd1}) and
(\ref{NPf}) we get
\be
{\ph}(x_1)\dots\ph(x_{N+1})=
\mbox{N}\left[\left.\PPr_{N+1}
\vp_1\dots\vp_{N+1}\right|_{\dots}\right]\ .
\nonumber
\ee
Hence the proof of Theorem~1 is completed. \
$\framebox(6,6){\ }$
\vspace{0.2cm}

{\bf Corollary 1.1}
Due to the linearity of the expression (\ref{normformr}) on each
variable
$\vp_i$ the formulae \re{normformr} can be rewritten in the following
form:
\be
\ph(x_1)\dots\ph(x_n) = \N \left[\left.
\exp\left[\sum_{i<k}\derr{\vp_i}n\derr{\vp_k}\right]
\vp_1(x_1)\dots\vp_n(x_n)\right|_{\dots}\
\right]
\lab{nfr}
\ee
Below we will use this relation to find the compact functional form for
the Wick's theorems for the symmetrical and chronological products.
\vspace{0.2cm}

{\bf Corollary 1.2} Formulae (\ref{normformr}, \ref{nfr}) hold true if
we
substitute right derivatives by left ones, sign ``$<$" in product limits
by
``$>$" and normal contraction $n$ by $n^{\mbox{\tiny T}}$.

Let us note here that the formal form of the statement of Theorem~1
does not depend on any statistics ({\it i.e.} is the same as in fermionic and
bosonic cases) and the commutation factors are ``hidden" in the nature
of the classical variables. It is useful to remember this when studying all
other theorems of the paper where it is also true.

\section{Wick's theorem for $q$-symmetrical products}

In this section we consider other kinds of products, in particular,
$q$-symmetrical and $q$-chronological. Such products naturally play the
role of the usual symmetrical for bosons (antisymmetrical for fermions)
and chronological products in the framework of the standard quantum
field theory. Moreover they are interesting by themselves because they
keep the symmetry properties of the $q$-operators.

In close analogy with the undeformed cases we define $q$-Symmetrical
product ($\Sym_q$-product) of $q$-operators in the following way
\be
\Sym_q[\ph(x_1)\dots\ph(x_n)] = \frac{1}{n!}\sum_{P}Q_{P}
P[\ph(x_1)\dots\ph(x_n)]\ .
\lab{sym-q-def}
\ee
The sum is taken over all $n!$ permutations $P$ of the $q$-operators
$\ph$ with corresponding statistical phases $Q_P$. These $Q_P$ are the
factors arising from permutation of operators from the original
order to the order $P$. We assume that under the permutation
$q$-operators $\ph$ are replaced by the corresponding classical
variables
$\ph\rightarrow\vp$ and we do not pick up the expressions due to the RHS
of the eq.(\ref{qphiop}). In other words $Q_P$ is defined from the
relation
\be
\vp(x_1)\dots\vp_n(x_n) = Q_P P[\vp(x_1)\dots\vp_n(x_n)]\ .
\lab{Q-P}
\ee
We now consider the $q$-chronological product. We assume that
$q$-operators $\ph$ depend also on time $t$ and define $q$-chronological
product (or $\T_q$-product) of the $q$-operators by the equation (all
times $t_i$ are assumed to be different):
\be
\T_q[\ph(x_1)\dots\ph(x_n)] = \sum_{P}Q_{P}
P[\theta(1\dots n)\ph(x_1)\dots\ph(x_n)]\ ,
\lab{T-q-def}
\ee
where  the $n$-point $\theta$-function is given by the relation:
\be
\theta(1\dots n)
\equiv \prod\limits_{k=1}^{n-1} \theta(t_k-t_{k+1}) \ .
\lab{teta}
\ee
The summation in eq.\re{T-q-def} is taken over all simultaneous
permutations of the operators $\ph(x_i)$ and the corresponding times
$t_i$ in \tet-function.

If all or any part of the arguments $t_i$ of the operators are
equal then the $\T_q$-product is not rigorously defined. We need to
complete the definition. We define $\T_q$-product under equal times
as the $\Sym_q$-product. An action of $\Sym_q$- and $\T_q$-products on
zeroth power of operators is defined as usual by the equalities:
$\Sym_q[1]=\T_q[1]=1$.

We would like to note that $\Sym_q$ and $\T_q$ (similar to $\N$) are not
true linear operator on the space of the field operators: an operator
equality $\hat{F}_1=\hat{F}_2$ is not followed by the equalities
$\T_q\hat{F}_1=\T_q\hat{F}_2$ or $\Sym_q\hat{F}_1=\Sym_q\hat{F}_2$
(because the field operators in the argument of operations
$\T_q,\Sym_q,\N$ behave themselves as the classical variables).

We call a product {\it $q$-symmetric} if under  permutation multipliers
($q$-operators) in the product it behaves in the same manner as
if the $q$-operators would be classical variables, {\it i.e.} only the
corresponding statistical phase appears (\ref{Q-P}). This definition
naturally generalizes the notion of (anti)symmetry of the field operator
products in the Bose (Fermi) case. For example, the normal product is
$q$-symmetric
\be
\N[\ph(x_1)\dots\ph(x_n)] =
Q_P\N[P\{\ph(x_1)\dots\ph(x_n)\}] \ ,
\lab{norm-sym}
\ee
It is also not
difficult to check that the $\Sym_q$- and
$\T_q$-products introduced above are $q$-symmetric:
\ba
{}\Sym_q[\ph(x_1)\dots\ph(x_n)]
&=& Q_P\Sym_q[P\{\ph(x_1)\dots\ph(x_n)\}] \ ,
\lab{sym-sym} \nonumber\\
{}\T_q[\ph(x_1)\dots\ph(x_n)] &=&
Q_P\T_q[P\{\ph(x_1)\dots\ph(x_n)\}] \ .
\lab{T-sym}
\ea
To formulate the functional form of the Wick's theorem for the
above defined products we need to introduce the $q$-chronological
contraction $\Dl_q(x,y)$ and the $q$-symmetrical contraction
(or the $q$-symmetric part of the normal contraction $n$) $n_s^q$.
They are defined by the relations
\ba
{}\T_q[\ph(x)\ph(y)] &=& {}\N[\ph(x)\ph(y)] + \Dl_q(x,y)\ ,
\lab{q-chron}
\\
{}\Sym_q[\ph(x_1)\ph(x_2)] &=&
{}\N[\ph(x_1)\ph(x_2)] + n_s^q(x_1,x_2)\ .
\lab{q-sym2}
\ea
Using definitions (\ref{normprod2}, \ref{sym-q-def}, \ref{T-q-def})
we get expressions for the contraction via normal one~(\ref{pairing}):
\ba
{\Dl}_q &=& \theta(12)n+\kappa Q\theta(21) n^{\mbox{\tiny T}}
= \theta(12)n(x_1,x_2)+\kappa Q(x_1,x_2) \theta(21)n(x_2,x_1)
\lab{Dl-q}
\\
n_s^q &=& \frac12 [n+\kp Q n^{\mbox{\tiny T}}]=
\frac12 [n(x_1,x_2)+\kp Q(x_1,x_2)n(x_2,x_1)]
\lab{n-s-q}
\ea
Due to our definition of the $q$-chronological product on equal times
we have
\be
\Dl_q(x_1,x_2)|_{t_1=t_2}=n_s^q(x_1,x_2)|_{t_1=t_2}
\lab{Dl-q-equal}\ .
\ee
We wish to note that the $q$-chronological contraction (similar to
$n_s^q$) in contrast to the normal contraction $n$ has the property of
$q$-sym\-met\-ry, {\it i.e.} $\Dl_q=\kp Q \Dl_q^{\mbox{\tiny T}}$.

We are now ready to formulate the Wick's theorem for $q$-symmetric
products which gives $\Sym_q$-product and $\T_q$-product in
the normal form.
\vspace{0.2cm}

{\bf Theorem 2}
\be
\Sym_q[\ph(x_1)\dots\ph(x_n)] = \N
\left[\left.
\exp\left(\frac{1}{2}\derr{\vp}n\derr{\vp}\right)
\vp(x_1)\dots\vp(x_n)\right|_{\vp=\ph}\
\right]
\lab{sym-qnorm}
\ee
\be
\T_q[\ph(x_1)\dots\ph(x_n)] =
\N\left[\left.
\exp\left(\derr{\vp}\Dl_q\derr{\vp}\right)
\vp(x_1)\dots\vp(x_n)\right|_{\vp=\ph}\right]
\lab{T-qnorm}
\ee
The proof can be performed in close analogy with the undeformed case \ci{V1} and
the proof of Theorem~1.
\vspace{0.2cm}

{\bf Corollary 2.1}
The normal contraction $n$ in eq.\re{sym-qnorm} can be replaced by
$n_s^q$ from \re{n-s-q} due to the fact that the kernel of the
differential operation is automatically symmetrized:
$$
\der{\vp_i}n\der{\vp_k} =
\der{\vp_k}\kp Qn^{\mbox{\tiny T}}\der{\vp_i}\ ,\qquad
\der{\vp}n\der{\vp} = \der{\vp}n^q_s\der{\vp}\ .
$$

\section{Wick's theorems for $q$-operator functionals}

To describe the Green function technique and perturbation theory for the
systems of the $q$-deformed particles we also need the rules for
finding the normal form not only of the products of the operators but the
whole functionals as well. A $q$-operator expression $F(\ph)$ is said to
be {\it $q$-operator functional} if it has the following form:
\be F(\ph) =
\sum\limits_{n=0}^\infty \int\dots\int\!  {\rm d}x_1\dots{\rm
d}x_n\  F_n(x_1,\dots,x_n) \ph(x_1)\dots\ph(x_n)\ .  \lab{qopfunc}
\ee
The operator functional is completely defined by the set of its
(may be generalized and singular) coefficient functions $F_n(x_1\dots
x_n)$.  We will call an operator functional {\it $q$-symmetrical} if all
its coefficient functions satisfy the following relations:  \ba
F_n(\dots x_i\dots x_k\dots)&=&Q_{ik}^{-1}
F_n(\dots x_k\dots x_i\dots)
\lab{F-q-sym}
\\\dots{\vp}(x_i)\dots\vp(x_k)\dots&=&Q_{ik}
(\dots\vp(x_k)\dots\vp(x_i)\dots)
\nonumber
\ea
Classical functionals (which are obtained from the operator one by
replacing \ph\ by \vp) correspond to the $q$-symmetrical operator
functional. $q$-Symmetrical coefficient functions are uniquely
determined by the classical functional. Let's note also that any
$q$-symmetrical functional possesses the property $F(\ph)=\Sym_qF(\ph)$.
We call an operator functional $F(\ph)$ as {\it an operator functional
in normal form} if it possesses the property
$F(\ph)=\N[F(\ph)]$

We emphasize that operator functionals are defined just by their
coefficient functions (not by operator $F(\ph)$). The functions
$F_n$ determine operator $F(\ph)$ uniquely but the opposite
statement is not true in general.

The formulae \re{sym-qnorm} and \re{T-qnorm} are generalized
directly to operator functionals due to ``universality'' of the
reduction operation (we assume that operator functionals do
not contain time derivatives, though all considerations can be
extended to this case). Thus we obtain the following rules for
reducing operator functionals to the normal form which
we collect into Theorem~3:
\vspace{0.2cm}

{\bf Theorem 3}
\be \Sym_q
F(\ph) = \N \left.
\exp\left(\frac{1}{2}\derr{\vp}n\derr{\vp}\right)
F(\vp)\right|_{\vp=\ph}\
\lab{sym-q-func}
\ee
\be
\T_q F(\ph) =
\N\left.
\exp\left(\derr{\vp}\Dl_q\derr{\vp}\right)
F(\vp)\right|_{\vp=\ph}
\lab{T-q-func}
\ee
\vspace{0.2cm}

{\bf Corollary 3.1}
From these various formulae of the inverse and combined transformations
can be easily found. For example
\eqalph
\be
\N F(\ph) = \Sym_q
\left.
\exp\left(-\frac{1}{2}\derr{\vp}n\derr{\vp}\right)
F(\vp)\right|_{\vp=\ph}\
\ee
\be
\T_q F(\ph) = \Sym_q
\left.
\exp\left(\frac{1}{2}\derr{\vp}(\Dl_q-n)\derr{\vp}\right)
F(\vp)\right|_{\vp=\ph}\
\ee
\eqarab
We proceed with the consideration of the normal form of a product
of the operator functionals. We formulate the Wick's theorem
for product of operator functionals {\it in normal form}:
\vspace{0.2cm}

{\bf Theorem 4}
\be
\prod\limits_{k=1}^n \left[ F^{(k)}(\ph)\right] = \N\left\{\left.
\exp\left[\sum_{i<k}\derr{\vp_i}n\derr{\vp_k}\right]
\prod\limits_{k=1}^n F^{(k)}(\vp_k)\right|_{\dots}\
\right\}
\lab{nffunc}
\ee
From now on the non-commuting multipliers are believed to be
ordered in accordance with increasing index:
\be
\prod\limits_{k=1}^n \left[ F^{(k)}(\ph)\right] \equiv
F^{(1)}(\ph)\cdot\dots\cdot F^{(n)}(\ph)\ .
\ee

{\bf Corollary 4.1}
Starting from the basic Wick's theorem \re{normformr} and arguing
as was done in deriving eq.(\ref{sym-qnorm}) one can
obtain the following rule of reduction of a product of
{\it $q$-symmetrical functionals} to the normal form:
\be
\prod\limits_{k=1}^n \left[\Sym_q F^{(k)}(\ph)\right] =
\N\left\{\left.
\exp\left[\frac{1}{2}\sum_{i}\derr{\vp_i}n\derr{\vp_i} +
\sum_{i<k}\derr{\vp_i}n\derr{\vp_k}\right]
\prod\limits_{k=1}^n F^{(k)}(\vp_k)\right|_{\dots}\
\right\}
\lab{eq100}
\ee
\vspace{0.2cm}

{\bf Corollary 4.2}
The eqs.\re{eq100} are obviously generalized to the case when some
multipliers in the LHS are given in the N-form (not Sym${}_q$-form):
\be
\prod\limits_{k=1}^n \left[{\cal A} F^{(k)}(\ph)\right] =
\N\left\{\left.
\exp\left[\frac{1}{2}\sum_{\mbox{\small Sym}_q}
\derr{\vp_i}n\derr{\vp_i} +
\sum_{i<k}\derr{\vp_i}n\derr{\vp_k}\right]
\prod\limits_{k=1}^n F^{(k)}(\vp_k)\right|_{\dots}\ \right\}\ .
\lab{eq101}
\ee
Here $\cal A$ denote $\Sym_q$ or N and summation in the diagonal
terms of the quadratic form is only over functionals that stand in
$\Sym_q$-form.

If a product of operator functionals stands under the common sign
of some $q$-symmetri\-cal product ($\Sym_q$ or $\T_q$) it is
automatically $q$-symmetrized and one can use the usual formulae
\re{sym-q-func} and \re{T-q-func}. For example,
\be
\T_q
\left[\prod\limits_{k=1}^n F^{(k)}(\ph)\right] = \N\left.
\exp\left(\derr{\vp}\Dl_q\derr{\vp}\right)
\prod\limits_{k=1}^n F^{(k)}(\vp_k)\right|_{\vp=\ph}
\ee
In conclusion we note that it is possible to
complete the definition of the T$_q$-product not via the Sym$_q$-product but
by using the N-form.  This is equivalent to the definition of the
\tet-function entering into the chronological contraction under
coincided
times as follows:  $\tet(t_{12})|_{t_1=t_2}=0$. It is not difficult to
check that the formula (\ref{T-qnorm}) remain true
under this convention. The same can be said about formulae
(\ref{T-q-func}) but the classical functional entering
into this formulae should represent N-form of the corresponding
$q$-operator functional, {\it i.e.} $F(\ph)=\N F(\ph)$.

\section{Conclusion.}

In this paper we have formulated Wick's theorems for general $q$-operators
and $q$-operator functionals. The formal
expression does not depend on the statistics of the fields and the
statistical aspect is revealed only when the nature of the algebra of
the
corresponding classical variables is defined. This implies that
the functional approach to the Wick's theorems using the analog of the
Hori's
formulae is the most natural one. Moreover, the theorems of the paper show
that it is possible to take into account the exotic statistics keeping
the formal expression for the Hori's formulae and changing the
permutation relations for the classical analogues according to the rules
of
the statistics. This fact will allow us to derive the perturbation series
for
the theory and develop analogous constructions of standard quantum field
theory in $q$-functional form in a straightforward way.

\section*{Acknowledgments.}
We want to thank V.M.Agranovich for the drawing of our attention to the
problems of quantum optics where $q$-particles find deep applications.
We
are also grateful to A.S.Vasiliev for interesting discussions.
This work was supported by Grant of the Russian Fund of Fundamental
Investigations N 95-01-00548 and UK EPSRC grants GR/L29156 and GR/K68356.

\end{document}